\documentstyle[sprocl]{article}

\input psfig.sty

\def\Journal#1#2#3#4{{#1} {\bf #2}, #3 (#4)}

\def\NPB{{\em Nucl. Phys.} B}

\def\PRL{\em Phys. Rev. Lett.}
\def\PR{\em Phys. Rev.} 
\def\PRD{{\em Phys. Rev.} D}

\def\JPA{{\em J. Phys.} A}
\def\ZPC{{\em Z. Phys.} C}

\def\AP{\em Ann. Phys. (N.Y.)}
 
\def\NAT{\em Nature}
\def\CMP{\em Comm. Math. Phys.}

\def\DAN{\em Dokl. Akad. Nauk. SSR)}
\def\SPD{\em Sov. Phys. Dokl.}
\def\JETP{\em Sov. Phys.-JETP}
\def\ZETF{\em Zh. Eksp. Teor. Fiz.}
\def\UFN{\em Usp. Fiz. Nauk.}
\def\SPU{\em Sov. Phys.-Usp.}
\def\RMP{\em Rev. Mod. Phys.}
\def\AJP{\em Am. J. Phys.}

\newcommand {\be}[1]{
    \begin{eqnarray} \mbox{$\label{#1}$}  }
\newcommand{\ee}{\end{eqnarray}}
\newcommand{\pref}[1]{(\ref{#1})}

\newcommand{\beas}{\begin{eqnarray*}} 
\newcommand{\eeas}{\end{eqnarray*}}

\newcommand \half {\frac{1}{2}}
\newcommand\ie {{\it i.e. }}

\begin{document}

\title{UNRUH EFFECT IN STORAGE RINGS 
\footnote{Talk delivered at the 18th Advanced ICFA Beam Dynamics Workshop on 
Quantum Aspects of Beam Physics, Capri October 15-20, 2000.}
}
\author{JON MAGNE LEINAAS }

\address{Department of Physics \\ University of Oslo
\\ P.O. Box 1048, Blindern \\ N-0316 Oslo  \\ Norway}

\maketitle\abstracts{A uniformly accelerated system will get
thermally excited due to interactions with the vacuum fluctuations
of the quantum fields. This is the Unruh effect. Also a system
accelerated in a circular orbit will be heated, but in this case complications
arise relative to the linear case. An interesting question is in what sense the
real quantum effects for orbital and spin motion of a circulating electron can
be viewed as a demonstration of the Unruh effect. This question has been
studied and debated. I review some of the basic points concerning the
relation to the Unruh effect, and in particular look at how the electron can
be viewed as a thermometer or detector that probes thermal and other
properties of the vacuum state in the accelerated frame.  }

\section{Introduction}
A uniformly accelerated observer will -- at least in theory -- see Minkowski
vacuum as a thermally excited state, with a temperature determined by the
acceleration. Thus, in the accelerated frame a natural definition of the vacuum
state is the Rindler vacuum, which is different from the Minkowski vacuum\,\cite{Fulling73}.
When expressed in terms of the Rindler field quanta, Minkowski space is thermally
excited. A quantum system that is uniformly accelerated will
act as a detector or thermometer that probes the temperature of the Minkowski
vacuum state in the accelerated Rindler frame. When coupled weakly to the
radiation field, it will end up in a stationary state with a thermal
probability distribution over energy levels\,\cite{Davies75,Unruh76}.  

There is a close relation between
this effect -- the Unruh effect -- and the Hawking effect, \ie the effect that
a black hole emits thermal radiation\,\cite{Hawking74}. The accelerated observer
in Minkowski space then corresponds to a stationary observer with a fixed distance
from the event horizon of the black hole. This observer will detect the local
effects of the Hawking temperature\,\cite{Unruh76}.

An interesting question is whether the Unruh effect can be seen in any real
experiment. There have been several suggestions, but there are obvious problems
with the implementation of such experiments. A main problem is that the
acceleration has to be extremely high to give even a very modest temperature. Thus,
a temperature of $1K$ corresponds to an acceleration of $2.5 \cdot
10^{\,20}\,m/s^2$. A detector indeed has to be very robust to take such an
acceleration without being destroyed. 

Some years ago it was suggested by John Bell and myself that that an electron
could be viewed as such a detector\,\cite{Bell83}. It certainly is robust at this
level of acceleration and in an external magnetic field the measured occupation of
the spin energy levels would give a way to determine the temperature. However, for
linearly accelerated electrons the obtainable  temperature and the time scale for
reaching equilibrium, for realistic values of the accelerating fields, makes it
clear that such effects cannot not be seen in existing particle accelerators.
Recently there have been interesting suggestions of how to obtain much more
violent acceleration of electrons by use of laser
techniques\,\cite{Chen99,Chiao00}, but even in this case the time scale for spin
excitations is too long. (There are however suggestions of other ways to see
the Unruh effect in these cases\,\cite{Chen99}.)

For electrons circulating in a storage ring the situation is different.
The acceleration is larger than in linear accelerators and the time available is
sufficient to reach equilibrium. In the discussion with John Bell the question
came up: Could the the Sokolov-Ternov effect\,\cite{Sokolov63,Derbenev73,Baier71},
which predicts a equilibrium polarization lower than 100\% be related to the Unruh
effect in the sense that the upper spin energy level, measured in the rest frame,
is partly occupied due to the heating? In two papers this question was studied an
the answer was a qualified yes\,\cite{Bell83,Bell87}. (See also several later
papers, some of these included in the proceedings from the ICFA meeting in
Monterey two years ago\,\cite{Unruh99,Barber99,Leinaas99}.)  For circular motion
there are however important complications relative to the case of linear
acceleration. One point is that the simple connection to temperature is correct
only for linear acceleration. For circular motion the excitations may be
described in terms of an effective temperature, but this temperature is not
uniquely determined by the acceleration as in the linear case. However, as a more
important point, the electron does not act as a simple point detector. The spin
is also affected by oscillations in the orbit, and the Thomas precession makes
the direct coupling to the magnetic field act differently from the indirect
coupling mediated through oscillations in the particle orbit. The net effect is
that the correct expression for the electron polarization will deviate from one
derived by a naive application of the Unruh temperature formula.

In this talk I will give a brief review of the connection between the Unruh and the
Sokolov-Ternov effect and I will stress some points which I find interesting
concerning this connection. First I will review how the Unruh effect, in principle,
could be demonstrated as a spin effect for linearly accelerated electrons. Then I
will discuss the relation between linear acceleration and circular motion and
finally consider the spin effect for circulating electrons. I should stress
that all the basic elements in the discussion of the Unruh effect for electrons in a
storage ring have been presented in the papers referred to above. However, the
discussion of the spin effect I will do in a slightly different way by
treating the spin and orbital motion on equal footing.  

\section{Linear acceleration and the Unruh effect}
A pointlike object which is uniformly accelerated, \ie which has a constant
acceleration as measured in the instantaneous inertial rest frame, describes a
hyperbolic path in space time. We may associate a co-moving frame with
the motion, with the time like unit vector as tangent vector of the
trajectory and the three space like vectors spanning the hyperplane of
simultaneity as defined by the moving object. The directions of the three
(orthogonal) space like vectors are determined only up to a rotation, but
if a non-rotational frame is chosen (Fermi-Walker transported) this
degree of freedom is eliminated. The non-rotational frame is also a
stationary frame in the sense that the acceleration is fixed with respect
to the unit vectors. 

The local frame can be extended in a natural way to an accelerated
coordinate system, the Rindler coordinate system\,\cite{Rindler66}. This is a
stationary coordinate system in the sense that the metric is independent of the time
coordinate. The transformation from Rindler coordinates $(x',\tau)$ to Cartesian
coordinates $(x,t)$ is, with the $x$-axis chosen as the direction of
acceleration,
\be{Rindler}
x=(x'+\frac{c^2}{a})\cosh(\frac{a\tau}{c})\;, \;\;\;
t=(\frac{x'}{c}+\frac{c}{a})\sinh(\frac{a\tau}{c})\;, \;\;\;y=y'\;,
\;\;\;z=z'
\ee
The Rindler time $\tau$ is the proper time of the trajectory $x'=0\,\;
(y'=z'=0)\;$, which is assumed to be the trajectory of the pointlike object.
However any trajectory
$x'=const$ is equivalent to this in the sense of having a constant proper
acceleration, although with a value of the acceleration that depends on the
$x'$-coordinate.

The coordinate system $(x',\tau)$ is well behaved only in a part of space-time.
At finite distance from the object (here chosen as the origin) there is a
coordinate singularity. The two hyperplanes $x=\pm ct$ which intersect
there define event horizons for the accelerated object. For points with
$x<ct$ (behind the future horizon) an emitted light signal will not reach the
object at any future time. A light signal emitted from the object at any past
time will not be able to reach the points with $x<-ct$ (behind the past
horizon). 

The Rindler coordinate system is similar to the Schwarzschild coordinate system of
an eternal black hole. In fact the Rindler coordinate system can be seen as a limit
case of black hole geometry, for points at finite distance from the horizon with
the mass of the black hole tending to infinity. In this limit the space time
curvature vanishes and only the effect of acceleration (due to gravitation) remains.

The Rindler coordinate system can be regarded as constructed from a continuous
sequence of inertial frames, the instantaneous rest frames of the accelerated
object. This implies that the time evolution of a system, described in the accelerated
coordinate system, can be expressed in terms of the generators of the Poincar{\'e}
group, which transform between inertial frames for infinitesimaly different values of
$\tau$. For linear acceleration only the time translation and the boost generator
(in the
$x$-direction) are involved, and the Hamiltonian will have the form
\be{AccHam}
H'=H+\frac a c K_x
\ee
If $H'$ is independent of $\tau$, the situation is stationary in the accelerated
frame, and the ground state as well as the excited states of the accelerated system
can be defined as the eigenstates of $H'$. (Such a physical system has to have a
finite size to avoid problems related to the coordinate singularity.)

If a uniformly accelerated system, described by a stationary Hamiltonian $H'$, is
coupled to a quantum field and this field is in the Minkowski vacuum state, the
vacuum fluctuations will cause transitions in the accelerated system. The important
point is, as pointed out by Unruh\,\cite{Unruh76}, that transitions to higher
energy (with respect to
$H'$), is caused by positive frequency fluctuations with respect to the Rindler time
$\tau$. The quantum fields in the vacuum state have only negative frequency
components with respect to Minkowski time $t$, but in terms of $\tau$ they have both
positive and negative frequency parts.

Let us consider the case of an accelerated electron. Then $H$ and $K_x$ are
respectively the Dirac Hamiltonian and the boost operator of Dirac theory
for the inertial (rest) frame at time $\tau$. If we neglect the fluctuations
in the trajectory and simply constrain the particle coordinates to the
classical path $x'=0$, the Hamiltonian
$H'$ is reduced to a spin Hamiltonian of the form
\be{Hspin}
H_{spin}=\half \hbar (\vec{\omega}+\delta \vec{\omega}) \cdot \vec{\sigma}
\ee
with $\vec{\omega}$ determined by the external magnetic field and
$\delta\vec{\omega}$ by the radiation field,
\be{omega}
\vec{\omega}_0=-\frac{e}{2mc}g\vec{B'}_{ext} ,\;\;\delta \vec{\omega}=
-\frac{e}{2mc}g
 \vec{B'}_{rad}
\ee
(The primed fields refer to the inertial rest frame.) The simplest situation would
be to consider an external magnetic field
$\vec{B}_{ext}$ in the same direction (the $x$-direction) as the accelerating
(electric) field.

The coupling to the radiation field causes transitions between the spin up and
spin down states of the particle in the external magnetic field. Standard first
order perturbation theory gives for the transition probabilities per unit time

\be{int}
\Gamma _\pm =\left(\frac{eg}{4mc}\right)^2 \int\limits_{-\infty }^\infty 
\exp (\mp i\omega
\tau )C(\tau-i\epsilon )d\tau  
 \ee
with $\pm$ referring to transitions up/down in spin and with $C(\tau )$ as the
vacuum correlation function of the magnetic field along the accelerated trajectory,
\be{corr}
 C_B(\tau )&=&\left\langle {B'_\pm (\tau / 2)B'_\mp (-\tau / 2)} \right\rangle
\nonumber\\ 
&=&{{\hbar a^4} \over {2\pi c^7}}\left[ {\sinh \left( {{a \over
{2c}}\tau } \right)} \right]^{-4}\;,\;\;\;\;B'_\pm=B'_x\pm i B'_y,
\ee
 The correlation function
$C(\tau )$ has a periodicity property with respect to shifts in the imaginary
time direction,
\be{sym}
C_B(\tau +{{4\pi c} \over a}i)=C_B(\tau )
\ee
which makes it easy to solve the integral \pref{int} by closing the integration
contour along the shifted path $\tau +4\pi i\, c/a$. For the closed contour we
find\,\cite{Bell83}
\be{Gint}
\left( {1-\exp \left( {\pm 2\pi c{\omega  \over a}} \right)} \right)\Gamma
_\pm =\mp {{e^2g^2\hbar } \over {6m^2c^5}}\omega \,\left( {\omega ^2+{{a^2} \over
{c^2}}}
\right)
\ee
and the ratio between transitions up and down is
\be{R}
R={{\Gamma _+} \over {\Gamma _-}}=\exp \left( {-2\pi c{\omega  \over a}}
\right)
\ee

In an equilibrium situation the ratio between transitions up and down defines the
relative occupation probabilities of the two spin levels. The dependence of the
energy shows that it has the form of a Boltzmann factor corresponding to a
temperature
\be{Unruhtemp}
kT_U=\frac{a\hbar}{2\pi c}
\ee
This is the Unruh temperature, which is generally associated with an accelerated
system. It has exactly the same form as the temperature of a black hole,
\be{BHtemp}
kT_{BH}=\frac{\kappa\hbar}{2\pi c}
\ee
whith the acceleration $a$ corresponding to the the surface gravity, $\kappa$, of
the black hole.

The spin effect discussed above is a special realization of the Unruh effect. The
interesting point is that this effect has a universal character, it does not depend
on details of the accelerated system or on its coupling to the radiation field. The
temperature effect follows from general features of the vacuum state and special
properties of the accelerated trajectory which lead to the symmetry properties
\pref{sym} of the vacuum correlation functions. One should also note that the
limitation to point detectors is not a necessary restriction. For the case
considered here the fluctuations in the particle path can be taken into account.
The Hamiltonian $H'$ will then depend on both spin and orbital coordinates and the
electron in this sense has to be treated as an extended system. If the
accelerating fields give rise to a time independent Hamiltonian in the
accelerated frame, the probability distribution over the energy levels of
$H'$ will still have a thermal form. This follows from general symmetry
properties of the vacuum correlation functions in the Rindler coordinate 
system, and is closely related to
PCT-invariance\,\cite{Sewell82,Hughes85,Bell85}. An interesting
complication is that the (local) temperature of such an extended system
would vary over the extension of the system. On the other hand, such a
variation is also present for a hot system in a gravitational field,
where the variation in temperature is induced by the redshift which
follows from differences in the gravitational potential.

A system of linearly accelerated electrons could in principle be used to demonstrate
the Unruh effect, in the way discussed above. Thus, the spin polarization would depend
on the spin precession frequency as
\be{pol}
P(\omega)=\frac{1-R}{1+R}=\tanh\left(\pi c\frac{\omega}{a}\right)
\ee
and by varying the strength of the external magnetic field the functional form of $P$
could be demonstrated. Unfortunately, this is not a realistic situation for real
particle accelerators. If we use a value for the electric field
$E=10MV/m$ we find a rather low corresponding temperature
$T=0.7\cdot10^{-3}K$. However, the main complication is the long time for
reaching equilibrium. From the expression for the transition
probabilities we find a typical time of
$\tau=3\cdot10^{18}s$. In the lab frame this would be enhanced even
further by the time dilatation effect. So this is not very promising. If
much larger accelerations can be obtained the thermalization time $\tau$
would be strongly reduced (it varies with $a$ as
$a^{-3}$), but the limited time available in a linear accelerator is
nevertheless a serious problem.

The limitations present for linear accelerators are not there
for circular accelerators. Thus, much higher values for the proper
acceleration are obtained, mainly due to the relativistic $\gamma$ factor for the
transformation from the lab frame magnetic field to the rest frame electric
field. A typical acceleration (limited by synchrotron radiation effects) is
$a=3\cdot10^{23}m/s^2$ corresponding to a temperature of $T=1200K$. And more
importantly, for electrons in a storage ring the time needed to reach
equilibrium is available, typically of the order of minutes to hours. 

In the magnetic field of a circular accelerator, the electrons will gradually build
up a transverse polarization due to spin flip radiation. This is well-known from
calculations by Sokolov and Ternov\,\cite{Sokolov63}, and later by
others\,\cite{Derbenev73,Jackson76} and the effect has been seen in real
accelerators\cite{}. The polarization will under ideal conditions reach the
equilibrium value of 92\%.

\begin{figure}[t]
\centerline{\psfig{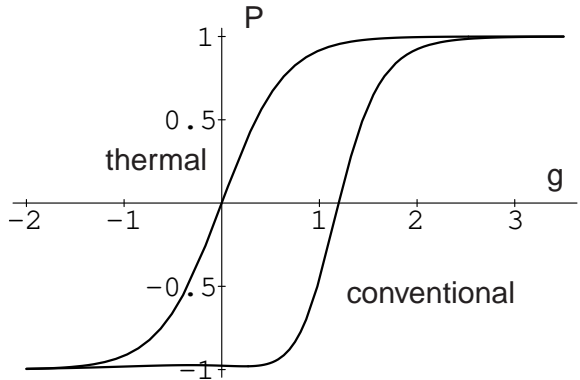}}
\vspace{15pt}
\caption{The equilibrium polarization $P$ as a function of $g$. The curve referred to
as "conventional" represents the conventional result, whereas the curve
referred to as "thermal" is based on the use of the relation between temperature
and acceleration derived in the case of linear acceleration.  \label{fig1}}
\end{figure}

In the figure
the theoretical curve for the polarization as a function of the $g$-factor is
shown. In the same figure also the corresponding curve, based on the simple formula
\pref{pol} obtained in the case of linear acceleration is shown. Much of the
discussion of the Unruh effect in storage rings has been based on comparison of
these curves. (For some critical remarks see Jackson\,\cite{Jackson99}.)
In the following I will examine this question again. There are two
important points involved in understanding the similarity and difference
between the two curves. The first is the question of the difference
between linear acceleration and acceleration in a circular orbit. The
other question concerns the approximation where we treat the electron as
a point detector.

\section{Stationary world lines}
The world line of a uniformly accelerated particle and the world line of a particle
moving with constant speed in a circular orbit are special cases of what has been
referred to as stationary world lines\,\cite{Letaw81}. These space-time curves are
self similar in the sense that there is no geometric difference between two
points on the trajectory. Such a curve can be generated by a time-independent
Poincar{\'e} transformation which in general will involve rotation in addition to
boost,\footnote{Such motion has been referred to as {\em group motion} and has been
examined in the context of relativistic {\em Born rigid
motion}\,\cite{Salzman54}.} 
\be{gen}
H'=H+{{\vec a} \over c}\cdot \vec K-\vec \omega \cdot \vec J
\ee
In the same way as for linear acceleration $H'$ can be seen as a time
evolution operator which jumps between inertial frames, the instantaneous rest frames
of the particle. It therefore generates a full (accelerated) coordinate system
in Minkowski space, a system where the particle sits at rest. This is a stationary
system in the sense that the space-time metric is independent of the time parameter.
The operator $H'$ can also be interpreted as the Hamiltonian of a quantum system
described in the accelerated frame. For the accelerated particle $H'$ is time
independent when the accelerating fields are stationary in this frame.

By making use of
the freedom to choose orientation of coordinate axis $H'$ can be brought into the form
\be{gen2} H'= H+{a
\over c}\;K_x-\omega _z\;J_z-\omega _x\;J_x
\ee
The physical interpretation of the parameters is that they correspond to acceleration
and angular velocity of a stationary frame moving with particle, as measured
relative to an inertial rest frame, but they also have a geometric
interpretation as curvature, torsion and hypertorsion of the world line of the
accelerated particle\,\cite{Letaw81}. 

The trajectory of uniform linear acceleration corresponds to $\omega_x=\omega_z=0$
and the circular orbit with constant acceleration and velocity corresponds to
$\omega_x=0, \, \omega_z=a/v$ with $v$ as the velocity of the particle. It is
interesting to note that a continuous interpolation can be made between the two cases
by changing $\omega_z$ while $\omega_x=0$. This looks as a purely formal
interpolation, since $v$ has to exceed the speed of light, but that is not really
the case. If the circular motion is not described in the rest frame of the center of
the orbit, but rather in the rest frame of the circulating particle at one of point
of the orbit, then there is a smooth interpolation between physically realizable
trajectories. For $\omega_z>a/c$, the trajectories in the $(x,y)$-plane are
cycloids, \ie they are periodic orbits. For $\omega_z\leq a/c$ they are
non-periodic (see figure.) The limit case $\omega_z=a/c$ can be interpreted as
the limit of circular motion where the radius  tend to infinity (and the velocity
$v\rightarrow c$) while the proper acceleration is fixed.
\begin{figure}[t]
\centerline{\psfig{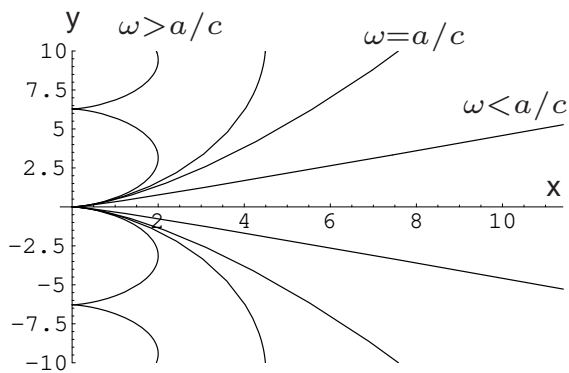}}
\vspace{15pt}
\caption{Stationary world lines projected on the $x,y$-plane for varying values
of the angular velocity $\omega$ with fixed proper acceleration $a$. 
\label{fig2}}
\end{figure}

In the same way as for linear acceleration all the acceleration dominated
trajectories $\omega_z<a/c$ have an event horizon. This horizon disappears in the
limit $\omega_z=a/c$ and is not present for the rotation dominated trajectories
$\omega_z>a/c$, \ie for circular motion. It is interesting to note the
similarity with the situation of a rotating (Kerr) black hole. The black hole is
characterized by two parameters, mass and spin (or surface gravity and angular
velocity). For fixed mass there is an upper limit to the angular velocity for
which the event horizon exists. For angular velocities beyond this one has the
(unphysical) situation with a space-time singularity but no event horizon. In
addition to the event horizon a rotating black hole is characterized by the
presence of a static limit, a limiting distance from the black hole, within which
no physical object can be stationary with respect to a distant observer. The
circular motion in Minkowski space corresponds to a situation with no event
horizon, but there is in this case a static limit outside which no object can be
stationary in the accelerated coordinate system. The presence of this limit is
easy to understand; due to the rotation points with fixed space coordinates will
at some distance from the center of rotation move with a velocity larger than the
velocity of light. At such a point a physical body cannot stay fixed.

As previously mentioned a detector which is uniformly accelerated through
Minkowski vacuum is related to a detector which sits at rest in a stationary
coordinate frame outside a (large) black hole. The black hole can be viewed as a
thermodynamic system, and in a stationary state the detector will be in thermal
equilibrium with the black hole, with a (locally determined) temperature that
depends on its position due to the redshift effect. This suggests that in a
similar way a detector following a trajectory with
$\omega_z\neq 0$ is related to a stationary detector outside a rotating
black hole. Also a rotating black hole can be viewed as a thermodynamic system,
with the angular velocity now acting as a chemical potential for the conserved
angular momentum. However, in this case we do not expect the occupation
probabilities over the detectors energy levels to be determined simply by the
temperature and angular velocity. This is because the detector will not be
coupled to the black hole in a rotationally invariant way. In a similar way
rotational invariance will be broken in the accelerated frame for any trajectory
with
$\omega_z\neq 0$, and there is no reason why the probability distribution in
this case should have a thermal form. Only for
$\omega_z=a/c$ that will be the case. The deviation from thermal form is
well-known for circular motion and also for the case $\omega_z<a/c$ this has
been discussed\,\cite{Letaw81}.

However, even though there is no exact temperature associated with the motion
where $\omega_z\neq 0$, the notion of an effective temperature is meaningful, as
has been discussed before. The vacuum state, in some approximate meaning seems
hot in the accelerated frame, and in addition there is a relative rotation between the
stationary detector and the vacuum. In the following I will re-examine the
case of the circulating electron from this point of view. The intention is to
show that such a picture of the vacuum state is relevant and to demonstrate that much
of the discrepancy between the polarization curve and the curve derived from the
assumption of thermal excitations is due to the
way the electron works as a detector.

\section{Electrons in a storage ring}
When we consider electrons in a storage ring under ideal conditions, where
they move in a rotationally symmetric magnetic field and with the correction
due to radiation loss neglected, they can be described by a
time-independent Hamiltonian $H'$ of the form previously discussed,
\be{Ham2}
H'= H+{a\over c}\;K_x-{a\over v}\;J_z
\ee
This Hamiltonian refers to an accelerated (rotating) coordinate system which rotates
with the frequency of electrons moving along a (classical) reference
trajectory in the magnetic field. The transformation between the accelerated and the
lab frame coordinates can be written as
\be{rotcoord}
x&=&\left( {x'+R} \right)\;\cos \left( {a\tau / v \gamma }
\right)-\gamma y'\;\sin \left( {a\tau / v \gamma } \right) \nonumber\\
y&=&\gamma y'\;\cos \left( {a\tau / v \gamma } \right)+\left( {x'+R}
\right)\;\sin \left( {a\tau / v \gamma } \right) \nonumber\\
z&=&z' \nonumber \\
t&=&\gamma \tau +\gamma v y'/c^2
\ee
with $R$ as the bending radius of orbit and $\gamma$ as the relativistic gamma factor.

The electrons described by the Hamiltonian are not restricted to
the reference trajectory $x'=y'=0$. Also quantum fluctuations
in the orbital motion are included. However, we may assume deviation from this
orbit both in position and velocity to be small, which makes it possible to linearize
in the deviation and to consider a non-relativistic approximation.

The operators $H$, $K_x$ and $J_z$ are Dirac operators in the inertial rest frame of
the reference trajectory at time $\tau=0$,
\be{HK}
H&=&c\vec {\alpha} \cdot \vec{\pi} + \beta mc^2
+e\phi+(g-2)\frac{e\hbar}{4mc}(i\beta \vec{\alpha} \cdot \vec{E}-\beta
\vec{\sigma} \cdot \vec{B}) \nonumber \\ 
K_x&=&-\frac{1}{2c} (xH +Hx) \nonumber \\
J_z&=&x p_y-yp_x +\half\hbar \sigma_z
\ee
A term for the anomalous magnetic moment $(g-2)$ has been introduced in the expression
for $H$.  All the coordinates and fields refer to the inertial rest frame of the
reference trajectory, but here and in the following I will omit the primes on these
coordinates and fields.
$\vec{\pi} =
\vec{p}-\frac{e}{c}
\vec{A}$ is the mechanical moment of the electron, and $\vec
{\alpha}$ and $\beta$ are the standard Dirac matrices.

When a Foldy-Wouthuysen transformation is performed and the equations are linearized
in the orbital fluctuations, the Hamiltonian can be reduced to a non-relativistic form
which involves  only spin and vertical oscillations. The coupling between the
horizontal oscillations and the spin can be considered as a higher order effect when
we primarily are interested in the spin degree of freedom. However the coupling to
vertical oscillations cannot be neglected, as has been discussed in earlier papers.
When we take into account only the external, accelerating fields the resulting electron
Hamiltonian gets the simple form
\be{Hame}
H_e={{p^2} \over {2m}}+{1 \over 2}m\Omega ^2z^2+{1 \over 2}\omega
\hbar
\sigma _z-{1 \over {mc}}\omega \hbar \,p\sigma _y-{\hbar  \over {4c}}g\Omega ^2z\sigma
_x\quad \quad 
  \ee
where $p$ is the momentum in the z-direction $\omega ={a \over {4c}}(g-2)$ is
the spin precession frequency. A confining harmonic oscillator potential has been
introduced for the vertical motion. In a weak focussing machine this is introduced by
a gradient in the magnetic field, $\Omega ^2=-{e \over m}{{\partial B} \over {\partial
r}}$.  The
coupling to the radiation field gives an additional term,
\be{H1}
H_1=-{{e\hbar } \over {4mc}}g\,\vec B\cdot \vec \sigma -e\,z\,E_z
\ee
which we treat as a perturbation. The fields $\vec B$ and $E_z$ are rest frame
fields.

Thus, the transformation to the accelerated frame gives us a fairly simple and straight
forward way to calculate the equilibrium properties of the electron beam. We first
determine the eigenvectors and eigenvalues of $H_e$, which is a two-level system
coupled to a harmonic oscillator, and we then find the occupation of the
levels by calculating the transition probabilities induced by $H_1$.

The coupling terms between spin and orbital motion in \pref{Hame} are normally
quite small. If they are treated perturbatively we find to first order the following
expressions for the eigenvectors of $H_e$
\be{evec}
\left| {\chi _{n,\pm }} \right\rangle =\left| {n,\pm } \right\rangle
-i\sqrt {{{\hbar \omega } \over {32mc^2}}}\left\{ {\left( {2\omega \mp g\Omega }
\right){{\sqrt {n+1}} \over {\omega \mp \Omega }}\left| {n+1,\mp } \right\rangle
} \right. \nonumber \\
 \left.
{+\,\left( {2\omega \pm g\Omega } \right){{\sqrt {n}} \over {\omega \pm
\Omega }}\left| {n-1,\mp } \right\rangle } \right\}
\ee
$\left| {n,\pm } \right\rangle$ is the eigenvector of the uncoupled system, with $n$
referring to the harmonic oscillator and $\pm$ to the spin levels. Note the resonance
between spin and orbital motion for
$\Omega=\omega$. The energy levels then are degenerate, and improved expressions close
to resonance could be found by use of degenerate perturbation theory, but I will not
do that here. Also note that to higher order in perturbation theory also higher order
resonances, for
$\Omega=n
\omega$, will be present.

Let us first consider the transition matrix elements for non-spin flip transitions.
It is clear from the form of $H_1$ that the small terms of \pref{evec} will only give
small contributions to the matrix elements. If they are neglected the result is
\be{ntrans}
\left\langle {\chi _{n+1,\pm }} \right|H_1\left| {\chi _{n,\pm }}
\right\rangle = ie\sqrt {{{\hbar \left( {n+1}
\right)} \over {2m\Omega }}}E_z
 \ee
For spin transitions the situation is different. Even if the coupling terms are small
they give significant contributions since they are influenced by the coupling of the
charge to the radiation field, which is much stronger than the spin coupling. The
result for spin flips is
\be{spintrans}
\left\langle {\chi _{n+1,\pm }} \right|H_1\left| {\chi _{n,\pm }} \right\rangle
=-{{e\hbar } \over {4mc}}\left( {gB_\mp +\left( {2-(g-2){{\Omega ^2} \over {\omega
^2-\Omega ^2}}} \right)E_z} \right)
\ee

The equilibrium populations $p_{n,\pm}$ of the energy levels can be found be assuming
detailed balance. Thus, the transition probabilities can be expressed in terms of the
electromagnetic fields, as for linear acceleration, and the relative populations can be
determined as the ratio between the probabilities for transition between
the levels one way and the other. Thus, for fixed spin the ratio is determined by the
matrix element \pref{ntrans}, and thereby by the correlation function of the
(rest frame) electric field,
\be{Romega}
R\left( \Omega  \right)&=&{{p_{n+1,\pm }} \over {p_{n,\pm }}}={{C_E\left(
\Omega  \right)} \over {C_E\left( {-\Omega } \right)}}, \\ \nonumber \\
C_E&=&\int {d\tau
}\,\exp \left( {-i\Omega \tau } \right)\,\left\langle {E_z\left(
\tau  \right)E_z\left( 0 \right)} \right\rangle
\ee
Note that the ratio $R(\Omega)$ is independent of the state $(n,\pm)$. This implies
that the excitation spectrum has a thermal form,
\be{pn}
p_{n,\pm }=N_\pm \exp \left( {-n\ln R} \right)
\ee
with a frequency dependent temperature\,\cite{Unruh99}
\be{efftemp}
eT_{eff}=\Omega \hbar\/ \ln R
\ee 

The relative population of spin up and down for the same $n$ is determined by ratios
between spin flip transitions. It has a similar form,
\be{spinflip}
R'\left( \omega  \right)={{p_{n,+}} \over {p_{n,-}}}={{D\left( \omega 
\right)}
\over {D\left( {-\omega } \right)}}
\ee
but now with a composite correlation function
\be{D}
D\left( \omega  \right)=g^2C_B\left( \tau  \right)+2gC_{EB}\left( \tau 
\right)+4C_E\left( \tau \right)
\ee
where $C_B$ is the correlation function of the magnetic fields $B_+$ and $B_-$ and
$C_{EB}$ is the mixed correlation function of $B_\pm$ and $E_z$.
\begin{figure}[t]
\centerline{\psfig{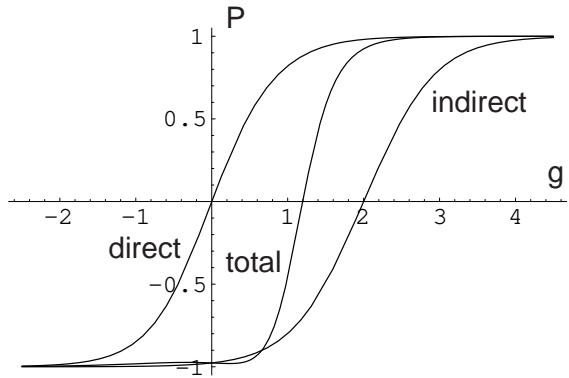}}
\vspace{15pt}
\caption{Evaluated polarization curves, based on Eq.\pref{spinflip} in the text.
"Total" refers to the full expression, and agrees with conventional result
displayed in Fig.\,1. "Direct" refers to the curve when fluctuations in the orbit
are suppressed, and "indirect" refers to the curve where only the spin excitations
due to fluctuations in the orbit are retained. 
\label{fig3}}
\end{figure}

The resonance term of
\pref{spintrans} is small except close to the resonance. If it is neglected, the
expression
\pref{spinflip} for the relative populations of the spin levels will reproduce the standard result
for the equilibrium polarization
$P=(1-R')/(1+R')$, earlier shown in Fig.1. It is interesting to see how this is
built up from contributions from the direct coupling between the spin and magnetic
field and the indirect one transmitted through fluctuations in the orbit. The
effect of the fluctuations is demonstrated by not including the two terms $C_{EB}$
and $C_E$ in
\pref{D}. The fluctuations in the orbit are then effectively suppressed and only
the coupling to the magnetic moment of the electron is retained. The result is a
changed curve $P(g)$, denoted "direct" in Fig.4. It indeed has a form very similar
to the one derived from the temperature formula, denoted "thermal" in Fig.1. The
effective temperature indicated by the curve is however somewhat higher than the
Unruh temperature $T_U$ for the same acceleration.

It is instructive also to consider the curve obtained if the contributions to the spin
transitions caused by the direct coupling is suppressed and only the indirect one, due
to fluctuations in the orbit, are retained. (That means including only $C_E$ in
Eq.\pref{D}.) The result is referred to as "indirect" in Fig.4. It is very similar
to the one obtained from the direct coupling to the radiation field, although
differing by a shift of 2 units along the g-axis.  

The relative shift of the curves obtained from the direct and indirect coupling of the
spin to the radiation field is fairly easy to understand.  The direct
coupling of the radiation field to the spin is rotationally invariant and the corresponding
occupation probabilities have (approximately) a thermal form in the frame which is
non-rotational (with respect to the vacuum). This is the Fermi-Walker transported frame,
where the direct coupling term is proportional to
$g$. The vertical fluctuations are insensitive to rotations in the
$x,y$-plane, but the coupling between the orbital motion and the spin is time
independent in the stationary frame, which is rotating relative to the Fermi-Walker
transported frame. This gives rise to occupation probabilities
which are (approximately) thermal in the stationary frame. The relative rotation
between these two frames is represented by  the shift along the $g$-axis.
\begin{figure}[t]
\centerline{\psfig{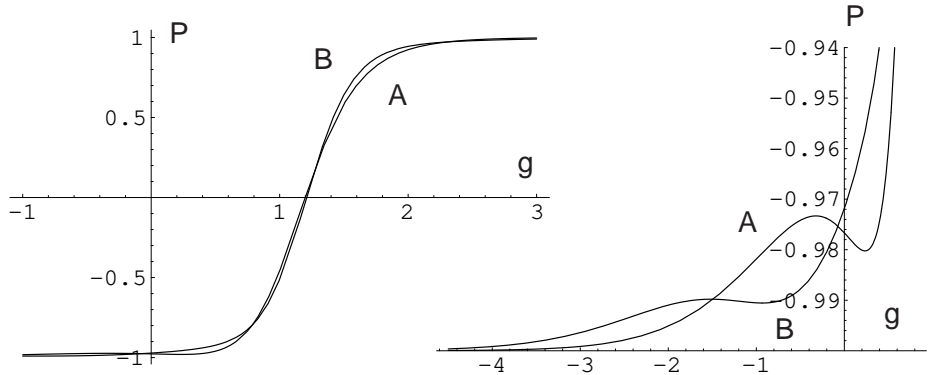}}
\vspace{15pt}
\caption{Comparizon of the conventional polarization curve (A) with the one
obtained by replacing the true correlation functions along the accelerated path
with thermal correlation functions (B). 
\label{fig4}}
\end{figure}

Based on this understanding we may conclude that much of the difference between the two
curves in Fig.1 is due to the way the electron acts as a detector. (See also the
related discussion by Unruh\,\cite{Unruh99}.) 
As a final point I will illustrate this by considering a hypothetical situation where the
detector, defined by the electron Hamiltonian \pref{Hame} and the coupling term \pref{H1},
is put in contact with an electromagnetic heat bath and where the detector is rotating
relative to the thermal state. Thus, the true correlation functions along the orbit are
replaced by thermal correlation functions in the evaluation of the polarization. In Fig.4 the
correct polarization curve is compared with the modified curve obtained in this way. The
temperature of the thermal state as well as the angular velocity are used as fitting
parameter. We note that a good approximation is obtained for an effective temperature
$T_{eff}\approx1.7 T_U$ and an angular velocity only slightly different from true
one, $\omega\approx1.1\;a/c$. There are some details, though, which are different
in the two cases, and which show that the correct curve is not truly thermal. 

\section{Concluding remarks}
An interesting observation made by Fulling\,\cite{Fulling73} and others is that
the natural definition of a vacuum state is related to the choice of the
space-time co-ordinates. In a curved space-time this makes the notion of a vacuum
state highly non-trivial and physical effects, like the Hawking radiation, may be
related to the question of what is the correct vacuum state.

Even in flat space, there are non-trivial vacuum effects associated with
accelerated co-ordinate frames. For a linearly accelerated system, the
Rindler coordinate system, Minkowski vacuum appears as a thermally excited
state. Also for other stationary coordinate systems, which include
rotation and not only acceleration, Minkowski vacuum appears as an excited
state, although not characterized by a uniquely defined temperature.

It is highly interesting that these vacuum effects, that usually are considered to
be detectable only under extreme situations, can be related to measurable
polarization effects of electrons in a storage ring. The natural way to see this
connection is to describe the electron beam in the accelerated frame of an
orbiting reference particle.The electron can be described as simple quantum
mechanical system, which includes the vertical motion coupled to the spin, and
with transitions between the states of this system induced by the radiation
field. Since the Minkowski vacuum state appears excited in this frame transitions
both up and down in energy are induced, and equilibrium is produced as a balance
between these two processes.

Due to the coupling between spin and orbital motion, the electron acts as a
non-trivial detector. Thus, the system is not rotationally invariant, and this
complicates the detection of the vacuum effects. The vacuum can be seen as being
hot in the accelerated frame, not in the stationary frame of the detector, but
rather the non-rotational (Fermi-Walker transported) frame. In fact, up to minor
details the correct polarization curve can be reproduced if the (rotating)
detector, defined by the electron Hamiltonian, is excited by an electromagnetic
heat bath, rather than the Minkowski vacuum state. 

The description of the orbiting electrons in the accelerated frame is
natural for the discussion of how the spin effects are related to the Unruh
effect. But I would also like to stress that it gives a conceptually simple
way to study the quantum effects of electrons in a storage rings. Under the
ideal conditions considered here, the electron is described as a harmonic
oscillator coupled to a two-level system with transition probabilities determined 
by correlation functions of the radiation field. It should be of interest to
examine this approach further, also beyond the simplest approximation
used here.

\end{document}